\documentclass[preprint,10pt,nofootinbib]{revtex4-1}
\usepackage[paper=letterpaper,margin=1.5in]{geometry}
\usepackage{amsmath,amssymb,amsfonts}
\usepackage{pstricks}
\usepackage{setspace}
\usepackage[varg]{txfonts}
\usepackage[off]{auto-pst-pdf}
\usepackage[]{graphicx}
\graphicspath{{graph/}%
}
\usepackage{epstopdf}
\usepackage[overlay]{textpos}
\usepackage{float}
\usepackage{hyperref}
\hypersetup{
  colorlinks,
  citecolor=green,
  linkcolor=blue}

\newcommand{\p}{\partial}
\newcommand{\const}{{\rm const}}

\renewcommand{\vec}[1]{\textnormal{\boldmath$#1$}}

\begin{document}
\normalcolor

\title{Short-range wakefields generated in the blowout regime of plasma-wakefield acceleration }

\author{G. Stupakov}
\affiliation{SLAC National Accelerator Laboratory,
Menlo Park, CA 94025}

\begin{abstract}

In the past, calculation of wakefields generated by an electron bunch propagating in a plasma has been carried out in linear approximation, where the plasma perturbation can be assumed small and plasma equations of motion linearized. This approximation breaks down in the blowout regime where a high-density electron driver expels plasma electrons from its path and creates a cavity void of electrons in its wake. In this paper, we develop a technique that allows to calculate short-range longitudinal and transverse wakes generated by a witness bunch being accelerated inside the cavity. Our results can be used for studies of the beam loading and the hosing instability of the witness bunch in PWFA and LWFA.
       
\end{abstract}

\maketitle


%
\section{Introduction}\label{sec:1}
%

In the blowout regime of both plasma-wakefield (PWFA) and laser-wakefield (LWFA) acceleration, plasma electrons are expelled from the region behind the electron bunch, or a laser driver, creating a cavity filled with the plasma ions~\cite{PWFA_blowout,Pukhov:2002rt}. The longitudinal electric field inside this cavity is used to accelerate a driven (or witness) electron bunch which is also being focused by the positive charge of the ions. The dynamics of the accelerating bunch can be strongly affected not only by the electromagnetic forces generated by the driver but also its own wakefields, both longitudinal~\cite{Tzoufras:2009yq} and transverse. In particular, a major concern for the future accelerators based on PWFA/LWFA is the beam break-up instability caused by the transverse wakefields in the cavity~\cite{Burov:2016fiw}.

To study the beam stability in traditional accelerators one usually invokes the concept of longitudinal and transverse wakefields~\cite{chao93}. They are calculated by solving Maxwell's equation with proper boundary conditions for two point charges---a leading source charge and a trailing witness charge---passing through the system, and then finding the longitudinal and transverse forces acting on the trailing charge. Given the wakefields for point charges, also  called the wake Green functions, one then calculates the forces inside the bunch for a given perturbation of the equilibrium  state and analyzes the stability of the perturbation. An important property of the wakefields used in this analysis is their linearity---the wake of a bunch is equal to the sum of the wakes generated by its charges. The wake linearity follows from the linearity of Maxwell's equations and the linearity of typical boundary conditions, such as at the vacuum-metal or vacuum-dielectric boundaries.

\begin{figure}[htb]
\centering
\includegraphics[width=0.48\textwidth, trim=0mm 0mm 0mm 0mm, clip]{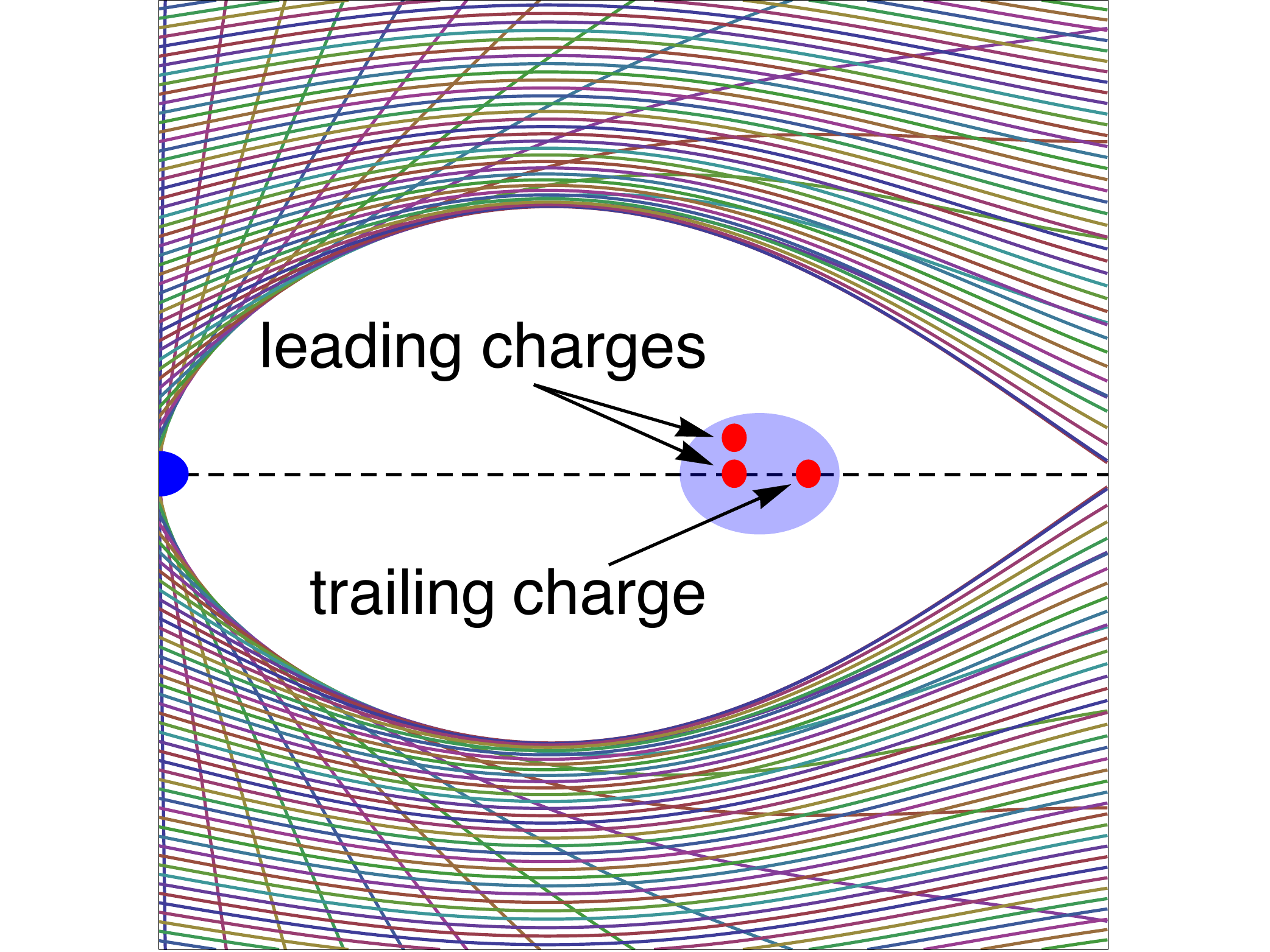}
\caption{Setup for calculation of the longitudinal and transverse wakefields in the blowout regime. The blue dot on the left is the driver. Trajectories of the plasma electrons are shown by colored lines. The light-blue blob is the driven bunch with bunch electrons shown by the red dots. For calculation of the longitudinal wakefield both the leading and trailing electrons are on the axis; for the transverse wakefield the leading electron is displaced from the axis.}
\label{fig:1}
\end{figure}
In this paper we extend the classical wakefield theory to the case of plasma acceleration in the blowout regime. The setup is shown in Fig.~\ref{fig:1}: an accelerating bunch is located inside an axisymmetric plasma cavity behind the driver. Following the general scheme of the wakefield theory, we consider two point charges of the accelerating bunch and calculate the longitudinal and transverse forces exerting by the leading charge on the trailing one. In the calculation of the longitudinal wakefield both charges are located on the axis of the system.  For the calculation of the transverse force we assume that the leading charge is displaced in the transverse direction.

In order for the plasma wakes to be useful for the standard stability analysis, they should be linear---only in this case the wake of a bunch of particles can be obtained by integration of the wake Green function with the bunch charge distribution. In general this is not true because the plasma wake is generated by a nonlinear flow of the plasma electrons responding to the presence of the witness bunch in the cavity. There are, however, two special cases when the linearity of the longitudinal wake holds. In the first case, the witness bunch charge is assumed so small that its field can be considered as a perturbation, and the equations describing the response of the plasma electrons to this perturbation can be linearized. In the linearized problem the wake is a linear function of the charges. In the second, more practically interesting case, instead of the smallness of the charge one assumes that the length of the witness bunch is small, $\sigma_z\ll k_p^{-1}$, where $k_p$ is the plasma wavenumber. The linearity of the wake in this case is not immediately evident---the remarkable property of the longitudinal wake linearity on short distances has been discovered in Ref.~\cite{barov_2004} for a bunch propagating in a uniform plasma. In this paper we extend the treatment of Ref.~\cite{barov_2004} for the calculation of the wake for  a short bunch inside the plasma cavity.

While the general method developed in this paper for the calculation of wakefields is applicable to both PWFA and LWFA, in our derivations in the subsequent sections we assume an electron driver bunch.

The paper is organized as follows. In Section~\ref{sec:2} we formulate equations for the steady-state plasma flow excited by a beam moving with the speed of light in a uniform plasma. In Section~\ref{sec:3} a concept of electromagnetic shock wave is introduced and is connected to the short-range longitudinal wake in a plasma cavity. In Section~\ref{sec:4} equations for the calculation of the longitudinal wake are derived. In Section~\ref{sec:5} the same procedure is applied to the transverse wakefield in the cavity. Numerical examples of wakefield calculations are presented in Section~\ref{sec:6}. Section~\ref{sec:7} summarized the results of the paper.

Throughout this paper we use the Gaussian system of units.

%
\section{Basic equations}\label{sec:2}
%

We assume that the driver and witness bunches move through the plasma with the speed of light $c$. Following the standard convention, throughout this paper we use dimensionless variables: time is normalized to $\omega_p^{-1}$, length to $k_p^{-1}$, velocities to the speed of light $c$, and momenta to $mc$. We also normalize fields to $mc\omega_p/e$, forces to $mc\omega_p$, potentials to $mc^2/e$, the charge density to $n_0e$, the plasma density to $n_0$, and the current density to $en_0c$. Here $e$ is the elementary charge, $e>0$. 

We consider a steady state in which all the quantities depend on $z$ and $t$ through the  combination $\xi= t-z$ (the $z$ coordinate is directed along the axis of the system in the direction of the beam motion). From the expressions for the fields in terms of the scalar potential $\phi$ and vector potential $\vec A$, $\vec E=-\nabla\phi-\p_\xi\vec A$, $\vec B    = \nabla \times \vec A$, it is easy to derive the following relations,
    \begin{align}\label{eq:1}
    \vec E_\perp
    =
    -\nabla_\perp\psi
    -
    \hat{\vec z}\times\vec B_\perp
    ,\qquad
    E_z
    =
    \p_\xi\psi
    ,
    \end{align}
where $\psi = \phi-A_z$, $\nabla = (\p_x,\p_y,-\p_\xi)$, $\hat{\vec z}$ is the unit vector in the $z$-direction, and the subscript $_\perp$ refers to the vector components perpendicular to the $z$ axis. Throughout this paper we use the notation $\p_a$ to denote differentiation with respect to variable $a$.

In our analysis, the plasma is treated as a cold fluid with ions represented as an immobile, positively charged neutralizing background. The equations of motion of plasma electrons are
    \begin{align}\label{eq:2}
    \frac{d\vec p}{dt}
    &=
    \nabla\psi
    +
    \hat {\vec z}\times \vec B_\perp
    -
    \vec v\times\vec B
    \,,
    \nonumber\\
    \frac{d\vec r}{dt}
    &=
    \frac{\vec p}{\gamma}
    \,,
    \end{align}
where $\vec p$ is the momentum, $\gamma$ is the Lorentz factor, $\gamma=\sqrt{1+p^2}$, and $\vec v$ is the velocity, $\vec v=\vec p/\gamma$. These equations conserve the quantity
    \begin{align}\label{eq:3}
    \gamma
    -
    p_z
    -
    \psi
    =
    \mathrm{const}
    =
    1
    ,
    \end{align}
where we have set the value of the constant equal to 1 because plasma electrons are initially at rest, $p_z=0$ and $\gamma=1$, in front of the driver where $\psi=0$.

In the steady state, the time derivatives in Eqs.~\eqref{eq:2} are replaced by the derivatives with respect to $\xi$, $d/dt = (1-v_z)^{-1}d/d\xi$. Using the relation $(1-v_z)^{-1} = \gamma/(1+\psi)$ that follows from Eq.~\eqref{eq:3}, the equations of motion can be written as
    \begin{align}\label{eq:4}
    \frac{d\vec p_\perp}{d\xi}
    &=
    \frac{\gamma}{1+\psi}
    \nabla\psi
    +
    \hat {\vec z}\times \vec B_\perp
    -
    \frac{B_z}{1+\psi}
    \vec p_\perp\times\hat {\vec z}
    \,,
    \nonumber\\
    \frac{d\vec r_\perp}{d\xi}
    &=
    \frac{\vec p_\perp}{1+\psi}
    \,.
    \end{align}

The continuity equation for the plasma electrons be written as follows:
    \begin{align}\label{eq:5}
    \p_\xi [n_e(1-v_z)]
    +
    \nabla_\perp
    \cdot
    n_e\vec v_\perp
    =
    0
    ,
    \end{align}
where $n_e$ is the plasma electron density. In this equation, the electron flow is  given by the product $n_e\vec v$, which assumes that the velocity $\vec v$ is uniquely defined at each point.
This is not always true for the blowout regime: there may be several streams of flow at a given point in space with different velocities $\vec v_s$ and densities $n_{es}$; in this case the flow is given by the sum over all the streams, $\sum_s n_{es}\vec v_s$. Our consideration is also valid for this case, however, to simplify the notation, we will continue to write $n_e\vec v$ instead of the sum $\sum_s n_{es}\vec v_s$. Note that in our dimensionless variables, the plasma current $\vec j$ is equal to the electron flow with negative sign, $\vec j = -n_e\vec v$.

We now derive equations for the electromagnetic field. First, we substitute Eq.~\eqref{eq:1} into Gauss's law,
    \begin{align}\label{eq:6}
    \nabla
    \cdot
    \vec E
    =
    1-n_e
    ,
    \end{align}
and use Amp\`{e}re's law $\nabla\times \vec B = \p_\xi \vec E + \vec j$ to obtain
    \begin{align}\label{eq:7}
    \Delta_\perp
    \psi
    =
    n_e(1-v_z)-1
    .
    \end{align}
Differentiating Eq.~\eqref{eq:7} with respect to $\xi$ and using Eq.~\eqref{eq:5} one can also derive the following equation for the longitudinal electric field $E_z=\p_\xi\psi$, 
    \begin{align}\label{eq:8}
    \Delta_\perp
    E_z
    =
    \nabla_\perp\cdot\vec j_\perp
    .
    \end{align}
To obtain an equation for the magnetic field we take the curl of both sides of the Maxwell equation $\nabla\times \vec B=\p_\xi \vec E+\vec j$ which gives $\Delta_\perp\vec B = -\nabla\times \vec j$. The latter can be decomposed into the longitudinal and transverse parts,
    \begin{align}\label{eq:9}
    \Delta_\perp\vec B_z
    = 
    -
    \hat{\vec z}\cdot
    (\nabla_\perp\times\vec j_\perp)
    ,\qquad
    \Delta_\perp\vec B_\perp
    = 
    \hat{\vec z}
    \times
    \nabla_\perp j_z
    +
    \hat{\vec z}
    \times 
    \p_\xi\vec j_\perp
    .
    \end{align}
Sometimes it is convenient to use an equation that is obtained by projecting $\nabla\times \vec B=\p_\xi \vec E+\vec j$ onto the $z$-axis
    \begin{align}\label{eq:10}
    \p_\xi E_z
    +
    j_z
    =
    -\nabla_\perp
    \cdot
    (\hat{\vec z}\times \vec B_\perp)
    .
    \end{align}

It is important to emphasize here that at a given value of $\xi$ and with the known transverse distributions of the plasma density $n_e$ and the current $\vec j$, the transverse dependence of the fields are found uniquely from Eqs.~\eqref{eq:7} and~\eqref{eq:9}.


Eqs.~\eqref{eq:4}-\eqref{eq:9} are simplified in the case of axial symmetry when the three components of the electromagnetic field, $E_z = \p_\xi \psi$, $E_r=-\p_r\psi$ and $B_\theta$, together with $\psi$, depend on $\xi$ and $r = \sqrt{x^2+y^2}$. The equations of motion~\eqref{eq:4} reduces to
    \begin{align}\label{eq:11}
    \frac{d p_r}{d\xi}
    &=
    \frac{\gamma}{1+\psi}
    \p_r\psi
    -
    B_\theta
    \,\qquad
    \frac{dr}{d\xi}
    =
    \frac{p_r}{1+\psi}
    \,.
    \end{align}
Eq.~\eqref{eq:7} for the wake potential becomes
    \begin{align}\label{eq:12}
    \frac{1}{r}
    \frac{\p}{\p r}
    r
    \frac{\p}{\p r}
    \psi
    =
    n_e(1-v_z)-1
    ,
    \end{align}
and for the magnetic field we have
    \begin{align}\label{eq:13}
    \frac{\p}{\p r}
    \frac{1}{r}
    \frac{\p}{\p r}
    rB_\theta
    &=
    -
    \frac{\p}{\p \xi}
    n_ev_r
    -
    \frac{\p}{\p r}
    n_ev_z
    -
    \frac{\p n_d}{\p r}
    .
    \end{align}
Eq.~\eqref{eq:8} for $E_z$ can be integrated ones to give
    \begin{align}\label{eq:14}
    \frac{\p}{\p r}
    E_z
    =
    -
    n_ev_r
    .
    \end{align}
We will also use Eq.~\eqref{eq:10},
    \begin{align}\label{eq:15}
    \frac{\p E_z}{\p \xi}
    -
    n_ev_z
    =
    \frac{1}{r}
    \frac{\p}{\p r}
    rB_\theta
    .
    \end{align}

%
\section{Longitudinal wakefield inside the plasma cavity}\label{sec:3}
%

Before we consider a point charge moving inside a plasma cavity, we will briefly summarize the results of  the steady-state solution of a point charge $q$ propagating with the velocity of light in a plasma of uniform density~\cite{barov_2004,stupakov_BKS}. Let the charge position in the plasma be characterized by the coordinate $\xi=t-z=\xi_0$. Because of the assumption $v=c$, the plasma in front of the charge, $\xi<\xi_0$, is not perturbed. Behind the charge, at $\xi>\xi_0$, a flow of plasma develops with a sharp boundary between the region near the axis from where the plasma electrons are expelled and $n_e=0$, and the region at large radii, where the electron plasma density is not zero. The most salient feature of this solution is that the transition from $\xi<\xi_0$ to $\xi>\xi_0$ occurs through an infinitesimally thin layer in which the fields have a delta-function discontinuity,
    \begin{align}\label{eq:16}
    E_r
    =
    B_\theta
    =
    2\nu
    K_1(r)
    \delta(\xi-\xi_0)
    ,
    \end{align}
where $K_1$ is the modified Bessel function of the second kind, $\nu$ is the dimensionless magnitude of the charge, $\nu = qr_ek_p/e$, and $r_e=e^2/mc^2$ is the classical electron radius. In what follows, we will refer to this discontinuity as the \emph{electromagnetic shock wave}, or simply the \emph{shock}, for brevity.

We will now turn to the problem of wakefields excited by a point source charge $q$ located inside a plasma cavity at the longitudinal position $\xi=\xi_0$ behind the driver bunch. We assume an axisymmetric cavity and the source charge on the $z$ axis. Our goal will be to calculate the longitudinal wakefield immediately behind the charge, at $\xi=\xi_0^+$. We denote by $E_r(r,\xi)$, $E_z(r,\xi)$ and $B_\theta(r,\xi)$ the field generated by the driver beam only (that is in the limit $q=0$), and by $\tilde E_r(r,\xi)$, $\tilde E_z(r,\xi)$ and $\tilde B_\theta(r,\xi)$ the field in the presence of the witness point charge $q$. Again, from the causality, it follows that the field and the plasma flow in front of the charge $q$ do not change and hence $\vec E=\tilde{\vec E}$, $\vec B=\tilde{\vec B}$ in region $\xi<\xi_0$. The fields change behind the point charge, at $\xi>\xi_0$; moreover, we expect that there will be an electromagnetic shock wave, as described above, but with yet unknown radial distribution of the field amplitude in the shock. Hence, the general expression for the field with the source charge $q$ can be written as follows:
	\begin{subequations}\label{eq:17}
    \begin{align}\label{eq:17a}
    \tilde E_r
    =
    E_r(r,\xi)
    +
    \Delta E_r(r,\xi) h(\xi-\xi_0)
    -
    D(r,\xi_0)\delta(\xi-\xi_0)
    ,
    \\
    \tilde B_\theta\label{eq:17b}
    =
    B_\theta(r,\xi)
    +
    \Delta B_\theta(r,\xi) h(\xi-\xi_0)
    -
    D(r,\xi_0)\delta(\xi-\xi_0)
    ,
    \end{align}
    \end{subequations}
where $h(\xi)$ is the step function equal to one for positive arguments and zero otherwise, $\Delta E_r$ and $\Delta B_\theta$ denote the change of the field due to the charge $q$, and the terms with the delta function represent a shock wave with the radial profile $D(r,\xi_0)$. 
The longitudinal electric field $E_z$ exhibits a discontinuity, $\Delta E_z$, on the shock wave,
    \begin{align}\label{eq:18}
    \tilde E_z(r,\xi)
    =
    E_z(r,\xi)
    +
    \Delta E_z(r,\xi)
    h(\xi-\xi_0)
    .
    \end{align}
The jump $\Delta E_z(r,\xi_0)$ is the longitudinal wake generated by the charge $q$ immediately behind it; we will show that this field is proportional to the charge $q$.

When plasma electrons cross the shock wave at a given radius $r$, due to its infinitesimal thickness, their crossing time is vanishingly small. The crossing changes both the transverse and longitudinal velocities, however, one can neglect the transverse displacement of plasma electrons  and consider the crossing as occurring at a fixed value of $r$. The problem then reduces to the relativistic motion of plasma electrons in the field that depends only on one coordinate $\xi$. The solution to this problem is well known~\cite{landau_lifshitz_ecm,Hora_1989}; for the reader's convenience we reproduce it in Appendix~\ref{app:1}. The vector potential $\vec A$ in Appendix~\ref{app:1} for our case has only radial components $A_r$. Through the comparison of the expressions for the fields in the shock wave~\eqref{eq:A.1} with the last terms in Eqs.~\eqref{eq:17}, and the fact that $A_r\propto h(\xi-\xi_0)$, we find that the function $D(r,\xi_0)$ is equal to $A_r$ taken behind the shock, $D=A_r(\xi_0^+)$.   The solution~\eqref{eq:A.13}, \eqref{eq:A.14} shows that the shock generates a discontinuity in the radial flow $n_ev_r$:
    \begin{align}\label{eq:19}
    \Delta(n_e{v}_r)
    =
    \frac{ n_{e0}(r,\xi_0)}{ \gamma_0(r,\xi_0)}
    D(r,\xi_0)
    .
    \end{align}
Here $ n_{e0}(r,\xi_0)$ and $ \gamma_0(r,\xi_0)$ refer to the values in the unperturbed flow in front of the shock wave that are equal to the corresponding values in the absence of charge $q$.

We can now obtain an equation for the function $D(r,\xi_0)$. We substitute the magnetic field from Eq.~\eqref{eq:17b} into Eq.~\eqref{eq:13} and integrate it over $\xi$ from $\xi =\xi_0^-$ to $\xi=\xi_0^+$. The only contribution on the left-hand side comes from the delta-function term in Eq.~\eqref{eq:17b}, and on the right-hand side the integration of $\p_\xi( n_ev_r)$ gives a jump in $n_ev_r$. Using Eq.~\eqref{eq:19} for the jump we obtain a differential equation for $D(r,\xi_0)$,
    \begin{align}\label{eq:20}
    \frac{\p}{\p r}
    \frac{1}{r}
    \frac{\p}{\p r}
    rD(r,\xi_0)
    &=
    \frac{ n_{e0}(r,\xi_0)}{ \gamma_0(r,\xi_0)}
    D(r,\xi_0)
    .
    \end{align}
The variable $\xi_0$ is a parameter in this equation.

Substituting Eq.~\eqref{eq:18} into Eq.~\eqref{eq:15} and equating the terms proportional to $\delta(\xi-\xi_0)$ we relate $\Delta E_z$ to $D$:
    \begin{align}\label{eq:21}
    \Delta E_z(r,\xi_0)
    =
    -
    \frac{1}{r}
    \frac{\p}{\p r}
    rD(r,\xi_0)
    .
    \end{align}
Differentiating this equation with respect to $r$ and using Eq.~\eqref{eq:20} we can also obtain
    \begin{align}\label{eq:22}
    \frac{\p}{\p r}
    \Delta E_z(r,\xi_0)
    =
    -
    \frac{ n_{e0}(r,\xi_0)}{ \gamma_0(r,\xi_0)}
    D(r,\xi_0)
    ,
    \end{align}
which is more convenient in numerical calculations. Integration of this equation yeilds
    \begin{align}\label{eq:23}
    \Delta E_z(r,\xi_0)
    =
    \int_r^\infty
    dr'
    D(r',\xi_0)
    \frac{n_{e0}(r',\xi_0)}{ \gamma_0(r',\xi_0)}
    .
    \end{align}
Taken with the minus sign and normalized by the the dimensionless charge $\nu= qr_ek_p/e$, gives the longitudinal wake $w_l$ on the axis behind the source charge located at coordinate $\xi_0$ in the bubble, $w_l=\nu^{-1}\Delta E_z(0,\xi_0)$. In the next section, we will show how to calculate this  wake for given functions $n_{e0}$ and $\gamma_0$.

%
\section{Calculation of the longitudinal wake}\label{sec:4}
%

Before applying Eq.~\eqref{eq:23} to the wakefield inside the plasma cavity, we consider first two simpler examples of the calculation of the longitudinal wake with this equation.

In the first example, we consider a point charge $q$ moving with the speed of light through a plasma with uniform density, $n_e=1$. We can use all the results of the previous section by simply setting $\vec E=0$, $\vec B=0$, which means that there are no fields in front of the charge $q$. To find the structure of the shock wake we need to solve Eq.~\eqref{eq:20} with $\gamma_{0}=1$ and $n_{e0}=1$. The boundary conditions for the function $D(r)$ are the following: $D(r)\to 0$ when $r\to\infty$ and 
    \begin{align}\label{eq:24}
    D(r)\to \frac{2\nu}{r},
    \qquad\mathrm{when}\ \ 
    r\to 0.
    \end{align}
The second condition means that at small distances the plasma currents do not shield the vacuum magnetic field of the relativistic point charge, $B_{\theta,\mathrm{vac}} = 2\nu/r$. These two boundary conditions uniquely define the solution of Eq.~\eqref{eq:20}, $D=2\nu K_1(r)$, in agreement with Eq.~\eqref{eq:16}. Having found $D(r)$ we can now integrate Eq.~\eqref{eq:22} and find  $\Delta E_z$, $\Delta E_z = 2\nu K_0(r)$. Using the Taylor expansion of $K_0(r)$ for $r\ll 1$ we find that in this retion $\Delta E_z \approx -2\nu \ln(r)$: the longitudinal field behind the point charge has a singularity on the axis. The field at small $r$ is positive: it  accelerates plasma electrons in the direction $-z$, opposite to the direction the charge motion.

In the second example, a point charge propagates along the axis of a hollow cylindrical plasma channel of radius $a$. The plasma density is zero, $n_e=0$, for $r<a$, and $n_e=1$ for $r>a$. In vacuum, the right-hand sides of Eqs.~\eqref{eq:20} and~\eqref{eq:22} are zeros and solving these equations we find $\Delta E_z=\Delta E_{z0}=\const$ and $D = 2\nu/r-r\Delta E_{z0}/2$, where we have used the boundary condition~\eqref{eq:24} and also Eq.~\eqref{eq:21}. In plasma, at $r>a$, we have $D=C K_1(r)$ and $\Delta E_z=C K_0(r)$ where $C$ is a constant. The unknown constants $C$ and $\Delta E_{z0}$ are found from the continuity of $D$ and $\Delta E_z$ at $r=a$:
    \begin{align}\label{eq:25}
    \frac{2\nu}{a}
    -
    \frac{a}{2}
    \Delta E_{z0}
    =
    C K_1(a)
    ,\qquad
    \Delta E_{z0}
    =
    C
    K_0(a)
    .
    \end{align}
Solving these equations we obtain
    \begin{align}\label{eq:26}
    C
    =
    \frac{2\nu}{a}
    \left(K_1(a)+\frac{a}{2}K_0(a)\right)^{-1}
    =
    \frac{4\nu}{a^2K_2(a)}
    ,
    \end{align}
and the longitudinal wake
    \begin{align}\label{eq:27}
    w_l
    =
    \frac{4K_0(a)}{a^2K_2(a)}
.
    \end{align}
In the limit $a\gg 1$ we find that $w_l\approx 4/a^2$ which is a standard universal expression for the wake at a short distance behind a point charge propagating in a round pipe of radius $a$ with resistive, dielectric or corrugates walls~\cite{bane96s,baturin14}. In the opposite limit, $a\ll 1$, we find $w_l\approx 2\ln(2/a)-2\gamma_E$ with $\gamma_E=0.577$ the Euler constant. We see that in this example, as well as in the previous one, the electric field $\Delta E_z$ is proportional to the charge $\nu$. 

We now return to the problem of the longitudinal wakefield inside a bubble cavity in the blowout regime of PWFA. As in the previous example, inside the bubble at a given $\xi$, we have $\Delta E_z=\Delta E_{z0}=\const$ and $D = 2\nu/r-r\Delta E_{z0}/2$. Outside of the bubble we need to solve Eq.~\eqref{eq:20} with the boundary condition at infinity $D(r\to\infty)\to 0$. Let us denote by $\hat D(r)$ a particular solution of this equation that satisfies this boundary condition and also $\hat D(r_b)=1$, where $r_b$ is the radius of the cavity at the location $\xi_0$ of the source charge (in what follows, we omit $\xi_0$ from the arguments of all functions). The general solution at $r>r_b$ is $D(r) = C\hat D(r)$. Introducing the auxiliary function
    \begin{align}\label{eq:28}
    \Delta \hat E_z
    =
    \int_r^\infty
    dr'
    \hat D(r')
    \sum_s
    \frac{n_{e0}(r')}{\gamma_0(r')}
    ,
    \end{align}
we also have $\Delta E_z=C\Delta \hat E_z$. The unknown factor $C$ is found from the continuity of $D$ and $\Delta E_z$ analogous to Eqs.~\eqref{eq:25},
    \begin{align}\label{eq:29}
    C
    =
    \frac{2\nu}{r_b}
    \left(1+\frac{r_b}{2}\Delta \hat E_z(r_b)\right)^{-1}
    ,
    \end{align}
with the longitudinal wake on the axis $w_l =\Delta E_{z0}/\nu$ equal to
    \begin{align}\label{eq:30}
    w_l
    =
    \frac{2\hat E_z(r_b)}{r_b}
    \left(1+\frac{r_b}{2}\Delta \hat E_z(r_b)\right)^{-1}
    .
    \end{align}
For a given radial distributions of the plasma density $n_{e0}$ and the Lorentz factor $\gamma_0$ outside of the bubble the two functions $\hat D$ and $\Delta \hat E_z$ can be computed numerically at each cross section of the cavity. Examples of such numerical calculations are presented in Section....

%
\section{Transverse wake}\label{sec:5}
%

We will now calculate the dipole transverse wake. For this, we consider the charge $q$ as moving with an offset $\vec{a}$ inside  the plasma cavity, where the magnitude of $a$ is much smaller than $k_p^{-1}$. This charge can be considered as having a dipole moment $\vec{d}=q\vec{a}$. The electromagnetic field of a dipole moving with the speed of light in \emph{free space} is
    \begin{equation} \label{eq:31}
    \vec{E}=
    2\delta(\xi-\xi_0)
    \frac{2(\vec{d}\cdot\vec{r})\vec{r}-\vec{d} r^2}{r^{4}}
    \,,
    \qquad
    \vec{B}=\vec{\hat{z}}\times\vec{E}
    \,,
    \end{equation}
where $\vec r = (x,y)$ is the two-dimensional vector perpendicular to the $z$ axis and $\xi_0$  defines the longitudinal position of the dipole. Assuming that $\vec d$ is directed along $x$ we can write its electric field in the cylindrical coordinate system $r,\theta,z$ as
    \begin{equation} \label{eq:32}
    \begin{pmatrix} 
	{E}_r  \\
	{E}_\theta  
	\end{pmatrix}
    =
    \frac{2d}{r^2}
    \delta(\xi-\xi_0)
    \begin{pmatrix} 
	\cos\theta  \\
	\sin\theta  
	\end{pmatrix}
    \,.
    \end{equation}

When the relativistic dipole $\vec d$ propagates inside the plasma cavity it changes the cavity field at $\xi>\xi_0$, similar to the monopole case. Due to the vector nature of the dipole perturbation the monopole equations Eqs.~\eqref{eq:17} and~\eqref{eq:18} are now replaced by 
    \begin{align}\label{eq:33}
    \tilde{\vec E}_\perp
    &=
    \vec E_\perp
    +
    \Delta \vec E_\perp(x,y,\xi)
    h(\xi-\xi_0)
    -
    \vec D(x,y,\xi)\delta(\xi-\xi_0)
    \nonumber\\
    \tilde{\vec B}_\perp
    &=
    \vec B_\perp
    +
    \Delta \vec B_\perp(x,y,\xi)
    h(\xi-\xi_0)
    -
    \hat{\vec z}\times\vec D(x,y,\xi)\delta(\xi-\xi_0)
    \nonumber\\
    \hat E_z
    &=
    E_z
    +
    \Delta E_zh(\xi-\xi_0)
    ,
    \end{align}
where $\vec D$ is a two-dimensional vector, $\vec D = (D_x,D_y)$. Comparing the singular terms in Eq.~\eqref{eq:33} with the expressions for the fields~\eqref{eq:A.1}, and taking into account that $\vec A\propto h(\xi-\xi_0)$, we find that $\vec D = \vec A(\xi_0^+)$. The jumps in the transverse currents $nv_x$ and $nv_y$ are expressed through the vector potential using Eq.~\eqref{eq:A.13} which reduces to
    \begin{align}\label{eq:34}
    \Delta (n_e\vec v_\perp)
    =
    \frac{n_{e0}}{\gamma_0}
    \vec D
    .
    \end{align}
Eq.~\eqref{eq:20} is now replaced by
    \begin{align}\label{eq:35}
    \Delta_\perp
    \vec D
    =
    \Delta (n_e\vec v_\perp)
    =
    \frac{n_{e0}}{\gamma_0}
    \vec D
    .
    \end{align}
The equations for the change of the $z$-components of the electric field through the shock wave can be obtained from Eq.~\eqref{eq:10},
    \begin{align}\label{eq:36}
    \Delta E_z
    =
    -
    \nabla_\perp
    \cdot
    \vec D(x,y,\xi)
    ,
    \end{align} 
from which we obtained Eq.~\eqref{eq:21} in axisymmetric case.

Generalizing the vacuum expressions~\eqref{eq:32}, we now seek the function $\vec D$ in the following form:
    \begin{align}\label{eq:37}
    \begin{pmatrix} 
	D_r  \\
	D_\theta  
	\end{pmatrix}
    =
    u(r)
    \begin{pmatrix} 
	\cos\theta  \\
	\sin\theta  
	\end{pmatrix}
    \,.
    \end{align}
Substituting this equation into Eq.~\eqref{eq:35} we obtain a differential equation for the function $u(r)$,
    \begin{align}\label{eq:38}
    u''
    +\frac{u'}{r}
    -
    \frac{4 u}{r^2}
    =
    u
    \frac{n_{e0}}{\gamma_0}
    \,.
    \end{align}
Using Eq.~\eqref{eq:36} we find that $\Delta E_z=\Delta \hat E_z\cos \theta$ where
    \begin{align}\label{eq:39}
    \Delta \hat E_z
    =
    -
    \frac{\p u}{\p r}
    -
    \frac{2u}{r}
    \,.
    \end{align}
Inside the bubble the right-hand side of Eq.~\eqref{eq:38} is zero and the solution is $u=br^2+2d/r^2$ which gives for $\Delta E_z$
    \begin{align}\label{eq:40}
    \Delta \hat E_z
    =
    -4br
    ,
    \end{align} 
where $b$ is a constant that should be defined from the matching of $u$ at the bubble boundary, as it was done in the previous section. 

We can now find the transverse force using the Panofsky-Wenzel theorem~\cite{panofsky56w} that relates the longitudinal derivative of the transverse force with the transverse derivative of the longitudinal field (in our dimensionless units),
    \begin{align}\label{eq:41}
    \frac{\p \vec F_\perp}{\p \xi}
    =
    \nabla_\perp (\Delta E_z)
    .
    \end{align}
With account of Eq.~\eqref{eq:40} we can write $\Delta E_z$ as $\Delta E_z = -4br\cos\theta = -4bx$, from  which we conclude that the transverse force is in $x$-direction,
    \begin{align}\label{eq:42}
    F_x
    =
    -4b(\xi-\xi_0)
    .
    \end{align}
Because this expression was derived with an assumption that the charge offset was in the $x$ direction, it is clear that, in general, $ \vec F_\perp$ is directed along $\vec a$. The force $\vec F_\perp$ normalized by the charge $\nu$ and by the offset $a$ is the transverse wake inside the bubble.

In the next section we will present an example of numerical calculations of the wake inside a bubble cavity generated by a driver bunch.

%
\section{Numerical examples}\label{sec:6}
%

To be able to calculate the wakefields in the plasma bubble using the method developed in the preceding sections we first need to find the plasma flow behind the driver bunch without the charge that generates the wakefield. There are several established computer codes that can solve this problem  numerically~\cite{doi:10.1063/1.872134,HUANG2006658,PhysRevSTAB.6.061301}. They, however, are not ideal for our purpose, because the detailed profile of the plasma flow needed for the wakefield calculations is not readily available in the outputs of the codes. For this reason, we have developed a new, fast computational code PLEBS (stands for PLasma-Electron Beam Simulations)~\cite{PLEBS} implemented in a matlab script that solves the plasma flow behind the driver and can be run on a desktop computer. 

In our numerical example, the driver bunch has the following parameters: the rms bunch length $\sigma_z=13\ \mu$m, the rms transverse size $\sigma_r = 5\  \mu$m, the bunch charge 1 nC. The bunch  propagates in plasma with the density $n_0 = 4\times 10^{16}$ cm$^{-3}$ (the plasma length parameter is $k_p^{-1}=26\ \mu$m). These are parameters representative of the beam-plasma experiments at FACET-II National User Facility being constructed at SLAC~\cite{Yakimenko:2016lfe}.  

\begin{figure}[!htb]
\centering
\includegraphics[width=0.49\textwidth, trim=0mm 0mm 0mm 0mm, clip]{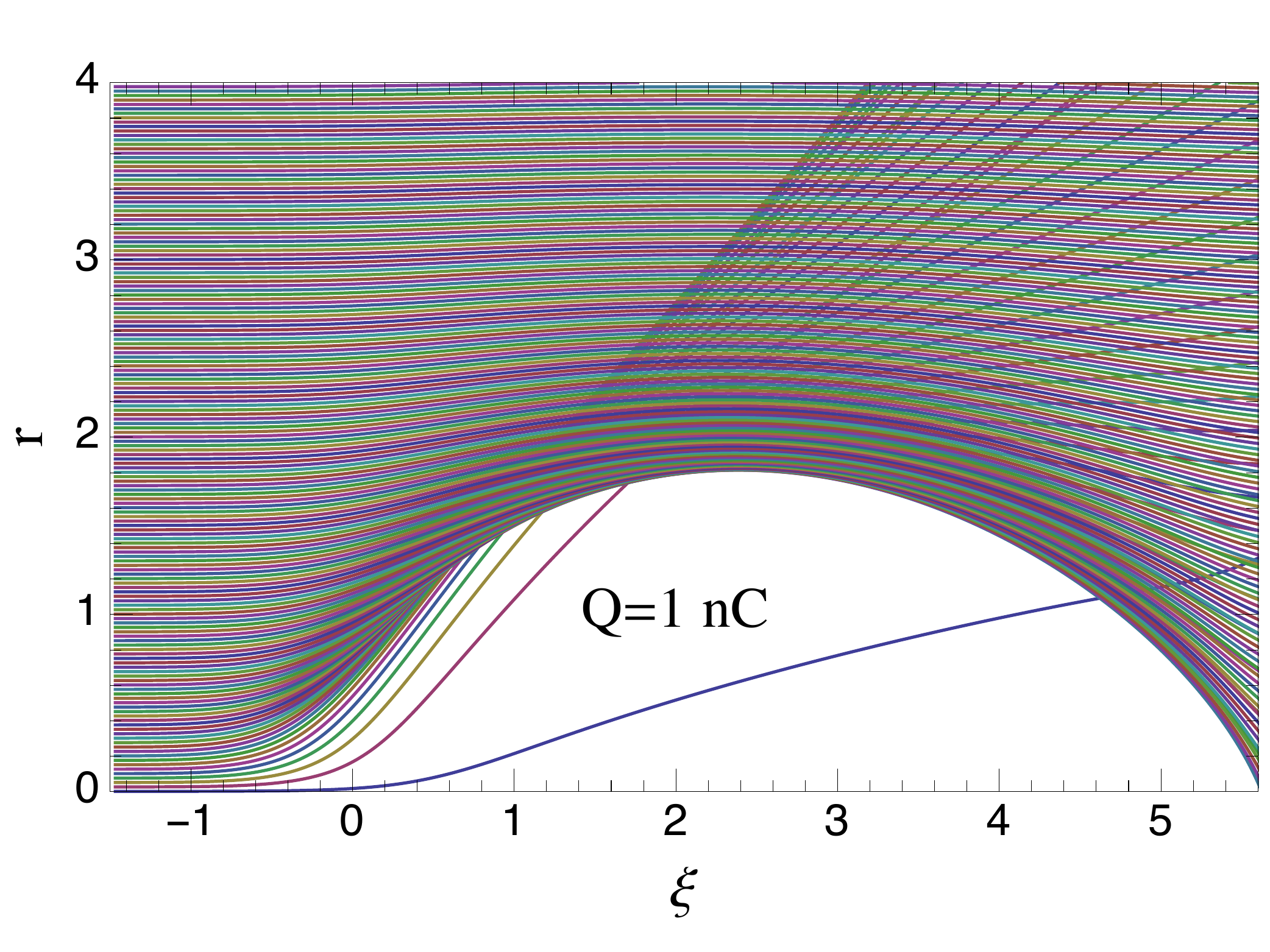}
\caption{Orbits of plasma electrons behind a driver bunch.  The center of the driver beam is located at $\xi=0$ and the rms driver length in dimensionless units is equal to 0.5.}
\label{fig:2}
\end{figure}

The driving beam has a peak current of 9 kA, and is dense enough to deflect plasma electrons passing through it to large radii and to form a plasma bubble shown in Fig.~\ref{fig:2}. Each line in this plot is a trajectory of a plasma electron. One can see a well defined boundary of the plasma bubble formed behind the driver. 

\begin{figure}[htb]
\centering
\includegraphics[width=0.49\textwidth, trim=0mm 0mm 0mm 0mm, clip]{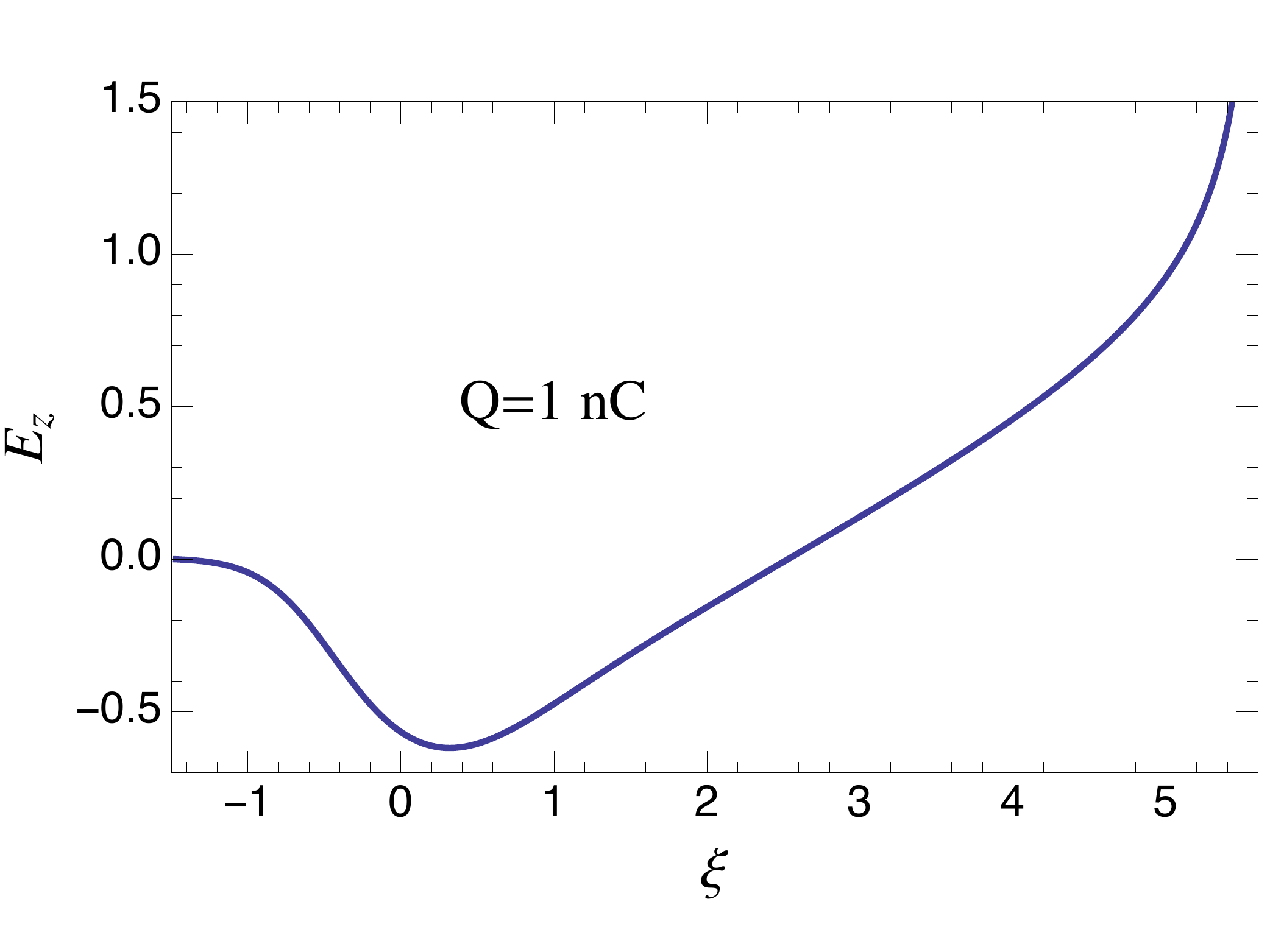}
\caption{The dimensionless longitudinal electric field on the axis of the bubble as a function of coordinate $\xi$.}
\label{fig:3}
\end{figure}

Fig.~\ref{fig:3} shows the longitudinal electric field on the axis of the bubble. Here, the negative values of $E_z$ correspond to the deceleration, and the positive values---to acceleration of electrons moving in the direction of the beam. One unit of the dimensionless electric field corresponds to the  gradient of 19.2 GeV/m.

\begin{figure}[htb]
\centering
\includegraphics[width=0.49\textwidth, trim=0mm 0mm 0mm 0mm, clip]{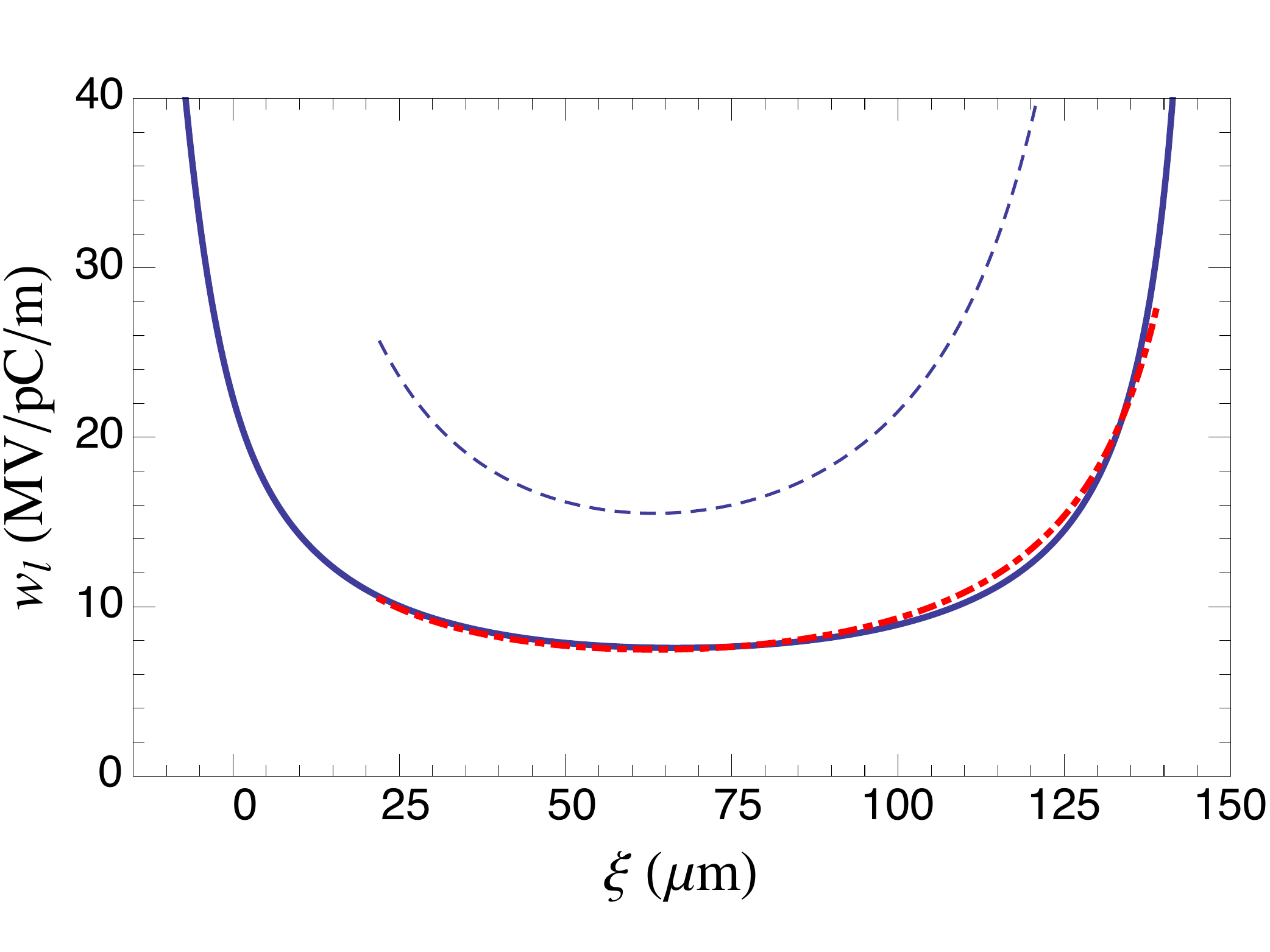}
\includegraphics[width=0.49\textwidth, trim=0mm 0mm 0mm 0mm, clip]{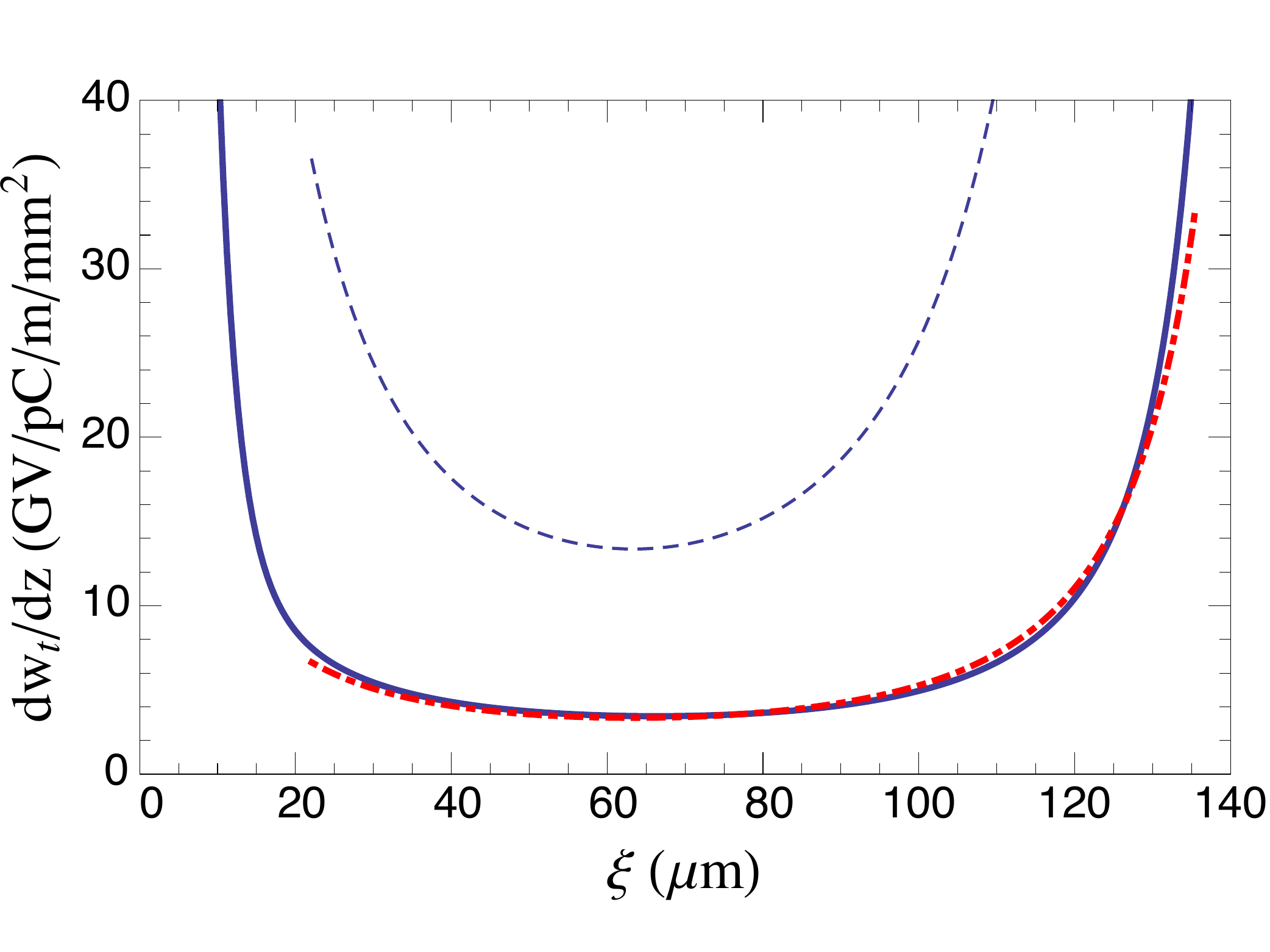}
\caption{Longitudinal wake (left panel) and the slope of the transverse wake (right panel) in the plasma bubble shown in Fig.~\ref{fig:2}. The dashed lines show the wakes calculated using the simple formulas for the short-range wakes in a cylindrical pipe. The red dot-dashed lines are plotted using Eqs.~\eqref{eq:43}.}
\label{fig:4}
\end{figure}

Using parameters of the plasma flow in this simulation we numerically solved Eqs.~\eqref{eq:20} and~\eqref{eq:38} for the strength of the shock wave for different positions $\xi$ of the source of the wakefield, and calculated the longitudinal and transverse wakefields. These wakes are shown by solid lines in Fig.~\ref{fig:4}.

It is interesting to compare the calculated short-range wakes with their analogs in a round pipe of radius $a$. It was already mentioned in Section~\ref{sec:4} that the longitudinal short-range wake is $w_l=4/a^2$, and it does not depend on the electrodynamic properties of the material wall of the pipe~\cite{bane96s,baturin14}. Similarly, the short-range transverse wake under the same conditions is a linear function of the distance $z$ between the source and the witness charges with the slope $d w_t/dz=8/a^4$. It was argued in Ref.~\cite{Burov:2016fiw} that these expressions for the wakes can be used as an approximation for the wakes in the limit of large charge of the driver when the pipe radius $a$ is replaced by the bubble radius $r_b(\xi)$ at the location of the source. These approximations are shown in Fig.~\ref{fig:4} by dashed lines---one can see that they overestimate the wake for our particular example of the bubble. The discrepancy between the simple formulas and our calculations can be explained in the following way. The electromagnetic field of the point charge $q$ penetrates into the plasma at distance of the order of $k_p^{-1}$ from the cavity boundary, which makes the effective bubble radius larger than $r_b$. Because $r_b$ enters into the formulas for the wake in the denominator, this makes the analytical estimate larger than the exact calculations. This explanation also offers a way to obtain more accurate estimates of the wake: replace the pipe radius $a$ by $r_b(\xi)+\alpha k_p^{-1}$, where $\alpha$ is a numerical coefficient of the order of one\footnote{The idea of fitting the plasma wakefields by formulas where $r_b$ is increased by $k_p^{-1}$ was proposed to the author by V. Lebedev.}. Our test of several cases show that a good approximation for the wakes is given by the following formulas
    \begin{align}\label{eq:43}
    w_l(\xi)=
    \frac{4}{(r_b(\xi)+0.8k_p^{-1})^2}
    ,\qquad
    \frac{d w_t}{dz}
    =
    \frac{8}{(r_b(\xi)+0.75k_p^{-1})^4}
    .
    \end{align}
Wakefields calculated using these formulas are also shown in Fig.~\ref{fig:4}, by dot-dashed red lines. We also simulated two more cases with the same dimensions of the bunch as above, but with different charges of the driver---2 and 4 nC.
\begin{figure}[htb]
\centering
\includegraphics[width=0.49\textwidth, trim=0mm 0mm 0mm 0mm, clip]{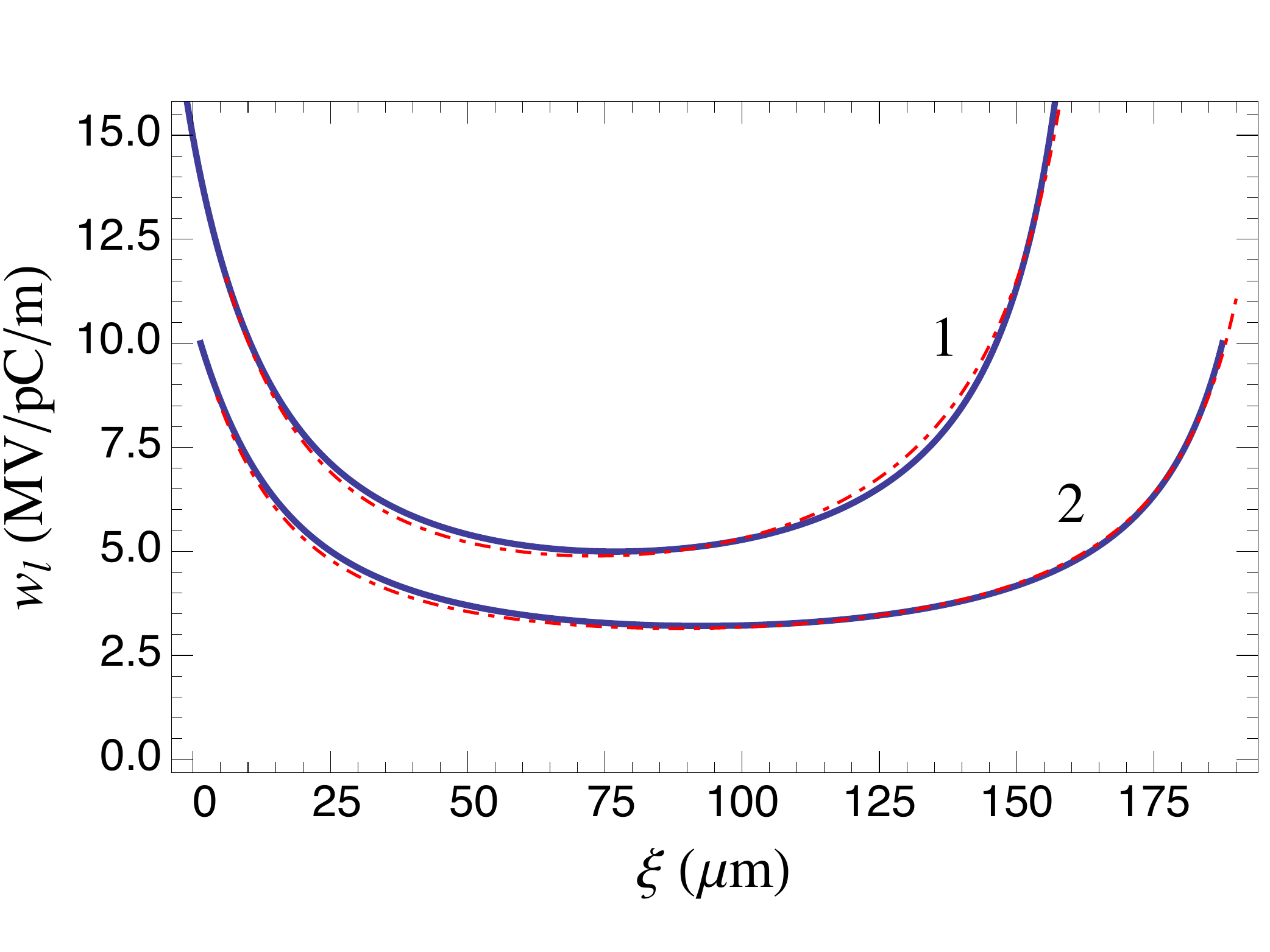}
\includegraphics[width=0.49\textwidth, trim=0mm 0mm 0mm 0mm, clip]{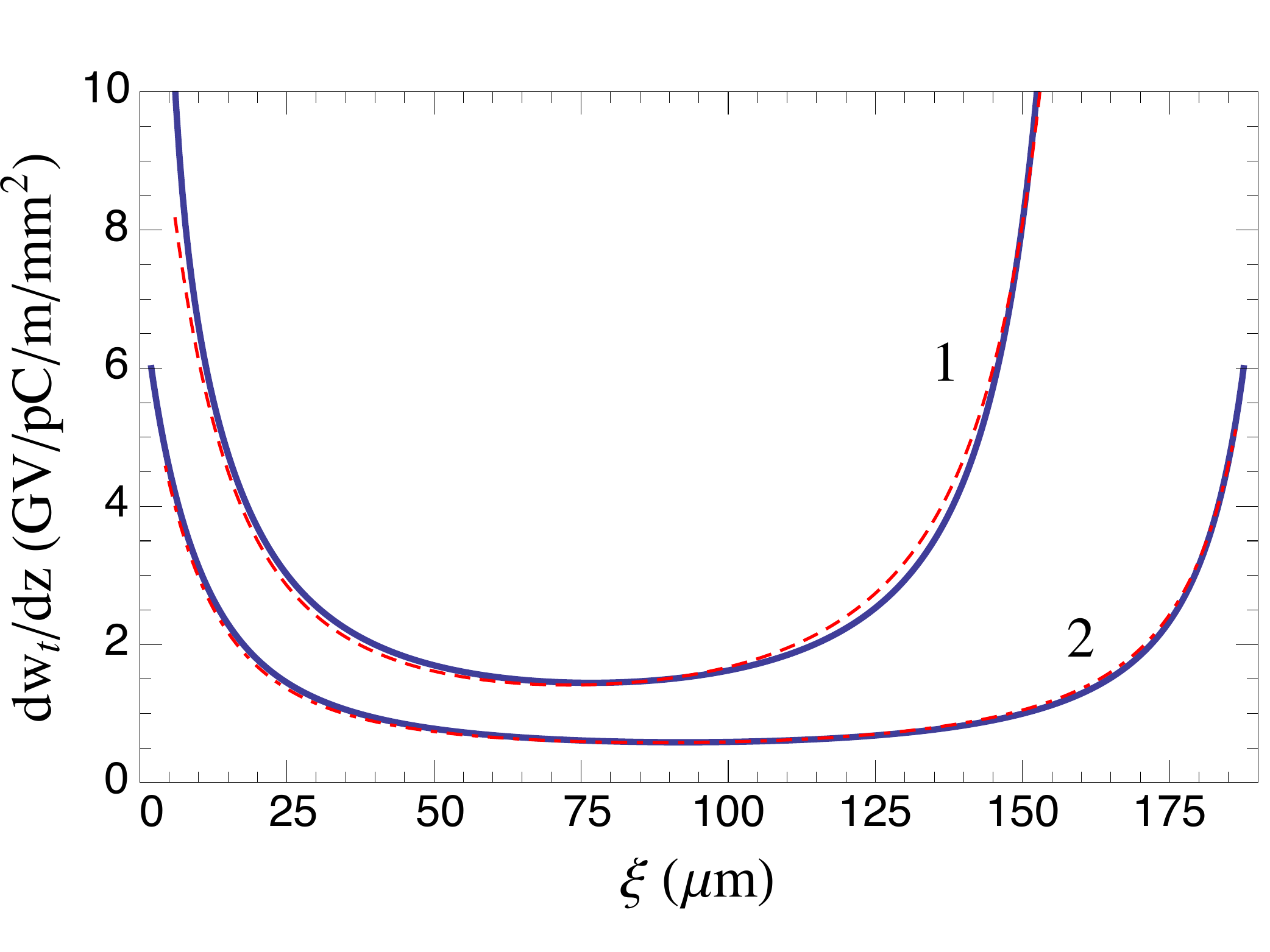}
\caption{Longitudinal wake (left panel) and the slope of the transverse wake (right panel) for 2 nC (label 1) and 4 nC (label 2) driver bunches. The red dot-dashed lines are plotted using Eqs.~\eqref{eq:43}.}
\label{fig:5}
\end{figure}
The calculated wakefields and the fitting formulas are shown in Fig.~\ref{fig:5}. We see that Eqs.~\eqref{eq:43} provide an excellent approximation for the wakes. Note also that a driver with a larger charge has smaller wakes---this is explained by the fact that a larger charge generates a bubble of a larger radius.

%
\section{Summary}\label{sec:7}
%

In this paper, we developed a method for calculation of short-range longitudinal and transverse wakefields in the blowout regime. We showed that these wakes scale linearly with the source charge, and they can be calculated from equations that involve the radial distributions of the plasma density and energy outside of the plasma bubble. We also showed  on several numerical examples that simple analytical formulas, Eqs.~\eqref{eq:43}, provide a good fitting to both longitudinal and transverse wakes.

%
\section{Acknowledgements}
%
The author thanks X. Xu for help with comparison of our results with QuickPic simulations and  P. Baxevanis for help with computer programming. The author is grateful to A. Burov for fruitful discussions.

This work was supported by the Department of Energy, contract DE-AC03-76SF00515. 

\newpage
\appendix

%
\section{Plasma Flow through Electromagnetic Shock Wave}\label{app:1}
%

As is explained in the main text, locally, we can neglect the transverse variation of the electromagnetic field in the shock wake. Directing the $x$-axis of the local coordinate system along the electric field in the shock, and the $y$-axis along the magnetic field, we express the field in the shock through the vector potential $\vec A=(A(\xi),0,0)$  such that
    \begin{align}\label{eq:A.1}
    E_x(\xi)
    =
    B_y(\xi)
    = 
    -A'(\xi)
    ,
    \end{align}
where the prime denotes the derivative with respect to the argument. The electric and magnetic fields in the shock wake are proportional to $\delta(\xi-\xi_0)$, which means that $A$ is the step function, $A(\xi) \propto h(\xi-\xi_0)$; however, in the derivation below we treat $A(\xi)$ as an arbitrary function of its argument.

Consider an electron located in front of the field where $A=0$ with an initial momentum $(p_{x0}, p_{y0}, p_{z0})$. As the electron begins to interact with the field its momentum changes according to the equation of motion,
    \begin{align}\label{eq:A.2}
    \frac{d\vec{p}}{dt}
    =
    -\vec{E}-\vec{v}\times \vec{B},
    \end{align}
from which we find that $d{p}_y/dt=0$ (we remind the reader that the negative sign on the right-hand side of this equation takes into account the negative electron charge). Hence $p_y$ is an integral of motion, $p_y = p_{y0} = \const$. For the motion in $x$-direction we have
    \begin{align}\label{eq:A.3}
    \frac{dp_x}{dt}
    =
    -E_x+v_zB_y
    =
    (1-v_z)A'
    =
    \frac{d}{dt}{A},
    \end{align}
where we have used the relation ${d}{A}/{dt} = (\p_t + v_z\p_z)A = (1-v_z)A'$. Hence $p_x-A = \const$ and taking into account the initial values $A=0$ and $p=p_{x0}$, we obtain
    \begin{align} \label{eq:A.4}
    p_x
    =
    A
    +
    p_{x0}.
    \end{align}

We now turn to the equation for $p_z$. We have
    \begin{align} \label{eq:A.5}
    \frac{d}{dt}{p_z}
    =
    -v_xB_y
    =
    v_xA'
    =
    \frac{1}{2\gamma}
    \left[
    (A^2)'
    +
    2A'p_{x0}
    \right]
    ,
    \end{align}
where we have used $v_x=(1/\gamma) (A+p_{x0})$. It is also useful to write down the equation for the $\gamma$-factor,
    \begin{align}\label{eq:A.6}
    \frac{d}{dt}{\gamma}
    = 
    -v_xE_x
    = 
    -v_xB_y
    =
    \frac{d}{dt}{p_z}
    \end{align}
from which it follows that $\gamma-{p_z}=\const =\gamma_0(1-v_{z0}),$ and 
    \begin{align}\label{eq:A.7}
    \gamma^{-1} 
    = 
    (1-v_z)\gamma_0^{-1}(1-v_{z0})^{-1}
    ,
    \end{align}
where $\gamma_0$  is the initial value of $\gamma$ before the acceleration. We will now show that the two terms on the right-hand side of Eq.~(\ref{eq:A.5}) are full time derivatives. For the first term, we have
    \begin{align}\label{eq:A.8}
    \frac{1}{\gamma}(A^2)'
    &= 
    \frac{1}{\gamma_0(1-v_{z0})}\left(1-v_z\right)(A^2)'
    = 
    \frac{1}{\gamma_0(1-v_{z0})}\frac{d A^2}{d t}.
    \end{align}
A similar transformation shows that the second term is proportional to $dA/dt$. This allows us to integrate Eq.~(\ref{eq:A.5}) over time:
    \begin{align}\label{eq:A.9}
    p_z 
    &=
    \frac{1}{2\gamma_0(1-v_{z0})}
    (
    A^2+2Ap_{x0}
    )
    +
    \gamma_0
    v_{z0}
    .
    \end{align}

Having solved a single particle motion through the shock wave, we will  now consider what happens to the plasma flow when it crosses the shock. Let $n_{e0}$ be the initial plasma density. Because we neglect the transverse derivatives, the continuity equation~\eqref{eq:5} reduces to $\p_\xi [n_e(1-v_z)]=0$, from which we find
    \begin{align}\label{eq:A.10}
    n_e(1-v_z)
    =
    n_{e0}(1-v_{z0})
    .
    \end{align}
As was mentioned in Section~\ref{sec:2}, there may be a multi-stream flow in the plasma---in this case Eq.~\eqref{eq:A.10} is valid for each stream.

We are also interested in the jump of the transverse flows  $n_ev_x$ and $n_ev_y$ through the shock wave. Using Eq.~\eqref{eq:A.4} we obtain
    \begin{align}\label{eq:A.11}
    n_ev_x
    =
    \frac{
    A
    +
    \gamma_0 v_{x0}
    }{\gamma(1-v_z)}
    n_{e0}
    (1-v_{z0})
    =
    n_{e0}
    v_{x0}
    +
    n_{e0}
    \frac{1}{\gamma_0}
    A
    ,
    \end{align}
where we have used~\eqref{eq:A.7}. We also have
    \begin{align}\label{eq:A.12}
    n_ev_y
    =
    n_{e0}
    \frac{1-v_{z0}}{\gamma(1-v_z)}
    \gamma_0v_{y0}
    =
    n_{e0}
    v_{y0}
    \end{align}
The last two equations can be combined into a vectorial one,
    \begin{align}\label{eq:A.13}
    n_e\vec{v}_\perp
    -
    n_{e0}\vec{v}_{\perp 0}
    =
    \frac{n_{e0}}{\gamma_0}
    \vec {A}
    ,
    \end{align}
where $\vec v_\perp$ and $\vec {A}$ are two-dimensional vectors with $\vec {A}$ defined by
    \begin{align}\label{eq:A.14}
    \vec A(\xi)
    =
    -
    \int_{-\infty}^\xi
    d\xi'
    \vec E_\perp(\xi')
    ,
    \end{align}
where $\vec E_\perp(\xi)$ is the transverse electric field in the shock wave. Eq.~\eqref{eq:A.13} is valid everywhere inside the shock--- we will use it for calculation of the jump in the transverse plasma flow through the shock wake taking the limit $\xi\to\infty$ in Eq.~\eqref{eq:A.14}.

\bibliography{\string~/gsfiles/Bibliography/master%
              }

\end{document}